\documentclass[12pt]{article}
\usepackage[margin=0.7in]{geometry}

\usepackage{amsmath}
\usepackage{amsfonts}
\usepackage{amsthm}
\usepackage{array}
\usepackage{multirow}
\usepackage{graphicx}
\usepackage{caption}
\usepackage{subcaption}
\usepackage{mathtools}
\usepackage{verbatim}

\newtheorem{theorem}{Theorem}

\newcommand{\entr}{s}
\newcommand{\vel}{u}
\newcommand{\temp}{\theta}
\newcommand{\dens}{\rho}

\newcommand{\press}{p}
\newcommand{\grav}{\mathrm{g}}
\newcommand{\PP}{\Phi}

\newcommand{\totalDiff}[1]{\frac{\mathrm{d}}{\mathrm{d}{#1}}}

\newcommand{\smbl}{\mathrm{smbl}}
\newcommand{\s}{\xi}

\newcommand{\systemEk}[1]{\mathcal{E}_{#1}}

\newcommand{\LieAlgebra}[1]{\mathfrak{#1}}
\newcommand{\Sdir}[1]{\partial_{#1}}

\title{Quotient of the Euler system on one class of curves}
\author{Anna Duyunova, 	Valentin Lychagin, 	Sergey Tychkov,
	\\ Institute of Control Sciences of RAS,
	\\ anna.duyunova@yahoo.com (A.D.), valentin.lychagin@uit.no (V.L),\\ sergey.lab06@yandex.ru (S.T.)
}
\begin{document}
\maketitle

\abstract{
	We consider the Euler system describing a one-dimensional inviscid flows in space along curves of a certain class.  
	Using differential invariants for the Euler system, we obtain its  quotient equation.
	The solutions of the quotient equation that are constant along characteristic vector field provide some solutions of the Euler system.
	We discuss solving the quotient using asymptotic expansions of unknown functions and virial expansion of thermodynamic state equations. Thus the quotient is reduced to a series of ODE systems.
}

\section{Introduction}
This paper is a continuation of the previous works (\cite{Duyunova2020_1,Duyunova2020_2,Duyunova2021_1,Duyunova2021_2, DLT2021}) on dynamics of gases and fluids on a space curve in a constant gravitational field. Before, we discussed symmetries and differential invariants of viscid \cite{Duyunova2020_2,Duyunova2021_1} and inviscid \cite{Duyunova2020_1} flows on space curves, and gave their classification based on symmetries group of the system.

In this paper, we continue our studies of one-dimensional inviscid flows in space along curves of a certain class. Specifically, using previously found differential invariants for the Euler system, we obtain its quotient. The latter, we use to find solutions of the Euler system itself.

Let us begin recalling \cite{Duyunova2020_1} that the system of PDEs describing flows along a space curve is the following
\begin{equation}\label{eq:Euler}
	\left\{
	\begin{aligned}
		&\dens(\vel_t  + \vel\vel_a)=- \press_a - \dens\grav h^{\prime},\\
		&\dens_t + (\dens\vel)_a=0,\\
		&\dens\temp\left(\entr_t + u\entr_a\right) - k \temp_{aa}=0,
	\end{aligned}
	\right.
\end{equation}
where $\vel(t, a)$ is the flow velocity, $\press(t, a)$, $\dens(t, a)$, $\entr(t, a)$,
$\temp(t, a)$ are the pressure, density, specific entropy, temperature of the fluid respectively,
$k$ is the constant thermal conductivity, $\grav$ is the gravitational acceleration,
$h(a)$ is the $z$-component of a naturally-parametrised space curve.

We are interested in the case when $h$ is quadratic, i.~e., the curve is $\{x=f(a),\,y=g(a),\,z=\lambda a^2\}$.

Earlier, in \cite{Duyunova2021_2, DLT2021}, we studied quotients of the Euler and Navier--Stokes systems given the function $h$ is linear.

The first thing that should be noted is that the system~\eqref{eq:Euler} is incomplete, namely, it lacks two additional relations between thermodynamic quantities. To obtain them, we employ the same method as we did in the paper \cite{DLTwisla}. 
The idea of this method is based on interpretation of media thermodynamic states as Legendrian, or Lagrangian, manifolds in contact, or symplectic, space correspondingly.

So, by the system  $\systemEk{}$ of differential equations, we mean the differential equations \eqref{eq:Euler} joint with equations of the thermodynamic state 
\begin{equation}\label{eq:Therm}
	L= \{\, F(\press,\dens,\entr,\temp)=0, \, G(\press,\dens,\entr,\temp)=0 \,\}
\end{equation}
that satisfy the relation 
\[
[F,G]=0 \quad \mathrm{mod} \quad \{F=0,\,G=0\} ,
\]
where $[F,G]$ is the Poisson bracket with respect to the symplectic form 
\[
\Omega = d\entr \wedge d\temp + {\dens^{-2}} d\dens \wedge d\press .
\]

The paper is organised as follows.

In Section \ref{sec:Euler}, we present the Euler system, and thermodynamic relations expressed in terms of the Planck potential. Then we reiterate results regarding symmetries and differential invariants for the case of $h(a)=\lambda a^2$.

In Section \ref{sec:Q}, the quotient is given, and some of its solutions are found. To this end, we calculate its symbol and characteristic vector fields. Thus we find a special class of solutions that are constant along characteristics.

In Section \ref{sec:virial}, we discuss solutions of the quotient \eqref{eq:QuotientEa^2} in a form of asymptotic expansion. This leads to a series of ordinary differential equations. Its zeroth term is a first-order equation, which we solve numerically and show its qualitative behaviour.

\section{Euler equations on a curve}\label{sec:Euler}

Consider the Euler system of differential equations on a space curve $\{x=f(a),\,y=g(a),\,z=\lambda a^2\}$ where constant $\lambda$ is positive:
\begin{equation}\label{eq:EulerH}
	\left\{
	\begin{aligned}
		&\dens(\vel_t  + \vel\vel_a)=- \press_a - 2\dens\grav\lambda a,\\
		&\dens_t + (\dens\vel)_a=0,\\
		&\dens\temp\left(\entr_t + \vel\entr_a\right) - k \temp_{aa}=0,
	\end{aligned}
	\right.
\end{equation}
here $\press$ and $\entr$ are expressed in terms of Planck potential $\PP(\dens,\temp)$ \cite{Lychagin2020}
\[
\press(\dens,\temp)=-R\dens^2\temp\PP_{\dens},\quad
\entr(\dens,\temp)=R(\PP+\temp\PP_{\temp}),
\]
and $R$ is a specific gas constant.

The symbol of the system \eqref{eq:EulerH} is 
\[
\smbl(E) = \begin{pmatrix}
\dens(\s_1+\vel\s_2)  & -R\dens\temp\s_2(\dens\PP_{\dens\dens}+2\PP_{\dens}) & -R\dens^2\s_2(\temp\PP_{\temp\dens}+\PP_{\dens}) \\
\dens\s_2 & \s_1+\vel\s_2 & 0  \\
0 & 0 & -k\s_2^2  
\end{pmatrix},
\]
and its determinant has the form
\[
\det \smbl(E) = -k\dens\s_2^2
\left((\s_1+\vel\s_2)^2 +R\dens\temp\s_2^2(\dens\PP_{\dens\dens}+2\PP_{\dens})  \right).
\]

Let us briefly recall the following results obtained in \cite{Duyunova2020_1}.
  
To describe the symmetry algebra of the Euler system $\systemEk{}$, we consider a Lie algebra $\LieAlgebra{g}$ of point symmetries of the PDE system \eqref{eq:EulerH}
\[
\begin{aligned} 
	&X_1 = \Sdir{t},\quad X_2 = \Sdir{\press},\quad X_3 = \Sdir{\entr},\quad X_4=\temp\,\Sdir{\temp},\\
	& X_5 = \press\,\Sdir{\press} + \dens\,\Sdir{\dens}-\entr\,\Sdir{\entr}, \qquad\,\,\,\,
	X_7 = \sin \omega t\,\Sdir{a} +  \omega \cos\omega t\,\Sdir{ u} ,\\
	&X_6 = a\, \Sdir{ a} +u\,\Sdir{ u}  -2 \dens\,\Sdir{ \dens} , \qquad 
	X_8 =\cos\omega t\,\Sdir{a} -  \omega \sin\omega t\,\Sdir{ u},	
\end{aligned} 
\]
where $\omega=\sqrt{2\lambda \grav}$.

Let $\vartheta \colon\LieAlgebra{g} \rightarrow \LieAlgebra{h}$ be the following Lie algebras homomorphism
\[
\vartheta\colon X\mapsto
X(\dens)\Sdir{\dens} + X(\entr)\Sdir{\entr} + X(\press)\Sdir{\press} + X(\temp)\Sdir{\temp},
\]
where $\LieAlgebra{h}$ is a Lie algebra generated by vector fields that act on the thermodynamic valuables $\press$, $\dens$, $\entr$ and $\temp$.
In our case, $\LieAlgebra{h}$ is generated by the vector fields 
\begin{equation*} 
	\begin{aligned} 
		&Y_1 = \Sdir{ \press}, \qquad  
		Y_{2} = \Sdir{ \entr} ,\qquad 
		Y_{3}=\temp\,\Sdir{ \temp}, \qquad 
		Y_{4} = \dens\,\Sdir{ \dens}, \qquad
		Y_{5} =\press\,\Sdir{ \press} - \entr\,\Sdir{ \entr} .   
	\end{aligned} 
\end{equation*}

Let also $\LieAlgebra{h_{t}}$ be the Lie subalgebra of the algebra $\LieAlgebra{h}$
that preserves thermodynamic state \eqref{eq:Therm}.

\begin{theorem}
	A Lie algebra $\LieAlgebra{g_{sym}}$ of symmetries of the Euler system $\systemEk{}$ coincides with 
	\[
	\vartheta^{-1}(\LieAlgebra{h_{t}}).
	\]
\end{theorem}

We consider two types of differential invariants of Euler system $\systemEk{}$ --- kinematic invariants (functions that are invariant with respect to the prolonged action of the Lie algebra $\LieAlgebra{g_m}$, which is the the kernel of the homomorphism $\vartheta$) and Euler invariants (with respect to the action of the Lie algebra $\vartheta^{-1}(\LieAlgebra{h_{t}})$).

\begin{theorem} \label{th-inv}
For the case $h(a)=\lambda a^2$, the kinematic invariants field is generated by the first-order basis differential invariants
	\[
	\dens,\quad
	\temp,\quad
	\vel_a,\quad
	\dens_a,\quad
	\temp_a,\quad
	\temp_t+\vel\temp_a
	\]
and by the basis invariant derivations
	\[
	\totalDiff{t}+\vel\totalDiff{a},\quad \totalDiff{a}. 
	\]
This field separates regular orbits.
The number of independent invariants of pure order $k$	is equal to $4$ for $k\geq 1$.
\end{theorem}

\section{Quotient equation}\label{sec:Q}
In this section we consider the quotient \cite{Duyunova2021_1} of the system \eqref{eq:EulerH} and some of its solutions.

For convenience, we choose $\dens$ and $\temp$ are the local coordinates on the quotient, hence we restrict our consideration to a domain, where
\begin{equation}\label{eq:indep}
\hat{d}\dens \wedge \hat{d}\temp \neq 0,\quad\text{or, equivalently,}\quad
\dens_a\temp_t-\dens_t\temp_a\neq 0,
\end{equation}
here $\hat{d}\dens$ and $\hat{d}\temp$ are total differentials of the invariants $\dens$ and $\temp$ \cite{Duyunova2021_1}.

It follows from Theorem \ref{th-inv} that there are four relations (syzygies) between the following second order invariants
\begin{equation}\label{eq:2ndcorder}
\frac{d u_a}{d\dens},\quad\frac{d \dens_a}{d\dens},\quad\frac{d \temp_a}{d\dens},\quad\frac{d(\temp_t + u \temp_a)}{d\dens},\quad
\frac{d u_a}{d\temp},\quad\frac{d \dens_a}{d\temp},\quad\frac{d \temp_a}{d\temp},\quad\frac{d(\temp_t + u \temp_a)}{d\temp},
\end{equation}
where 
\[
\frac{d}{d\dens} = \frac{1}{\dens_t\temp_a-\dens_a\temp_t} \left( \temp_a \totalDiff{t} - \temp_t \totalDiff{a} \right), \qquad
\frac{d}{d\temp} = \frac{1}{\dens_t\temp_a-\dens_a\temp_t} \left( - \dens_a \totalDiff{t} + \dens_t \totalDiff{a} \right)
\]
are Tresse derivatives.

Choosing $\dens$, $\temp$, $\vel_a$, $\dens_a$, $\temp_a$, $\temp_t+\vel\temp_a$ as Lie--Tresse coordinates $x$, $y$, $K$, $L$, $M$, $N$ respectively and eliminating second-order derivatives of functions $\vel$, $\dens$, $\temp$ from Tresse derivatives \eqref{eq:2ndcorder} due to the first prolongation of Euler system \eqref{eq:EulerH}, given
\[
\vel^2 - R\dens\temp(\dens\PP_{\dens\dens} + 2\PP_{\dens}) \neq 0,\quad
KMx+LN\neq 0, \quad
M\neq0,
\]
we obtain the quotient equation $E_q$ as the following PDE system for the functions $K$,$L$,$M$,$N$ of $(x,y)$.
\begin{equation}\label{eq:QuotientEa^2}
	\left\{
	\begin{aligned}
		&xKM_x-NM_y+LN_x+M(N_y-K)=0,\\
		&Rxy(xK(\PP_x+y\PP_{xy})-N(2\PP_y+y\PP_{yy}))+k(LM_x+MM_y)=0,\\
		&RL\left(\right. xy(\PP_{xxx}L^2+2\PP_{xxy}ML+\PP_{xyy}M^2)+
		(xyLL_x+xyML_y+2xLM+3yL^2)\PP_{xx}+ \\
		&(xyLM_x+M(xyM_y+2xM+3yL))\PP_{xy}+
		(2yLL_x+2yML_y+xLM_x+M(xM_y+3L))\PP_x\left. \right) +\\
		&xK^2L_x-KNL_y-(xKM+LN)K_y-3LK^2-\omega^2 L=0,\\
		&RMx\left(\right. xy(\PP_{xxx}L^2+2\PP_{xxy}ML+\PP_{xyy}M^2)+
		(xyLL_x+xyML_y+2xLM+3yL^2)\PP_{xx}+ \\
		&(xyLM_x+M(xyM_y+2xM+3yL))\PP_{xy}+
		(2yLL_x+2yML_y+xLM_x+M(xM_y+3L))\PP_x\left. \right) +\\
		&N^2L_y-xKNL_x+(xKM+LN)xK_x+2LKN-K^2x-\omega^2 x=0.
	\end{aligned}
	\right.
\end{equation}
Note that the condition $\vel^2 - R\dens\temp(\dens\PP_{\dens\dens} + 2\PP_{\dens})\neq 0$ is satisfied due to the negative-definiteness of the form $\kappa\vert_L$ \cite{DLTwisla}, and the condition $KMx+LN\neq 0$ is equivalent to \eqref{eq:indep}.

If $L\neq 0$, the last equation of the system above can be rewritten as
\[
x(MK_y-KL_x+LK_x)+NL_y+2KL=0.
\]
The symbol of the system \eqref{eq:QuotientEa^2}, denoted by $\smbl(E_q)$, is 
\[
\begin{pmatrix}
	0 & 0 & k(L\s_1+M\s_2) & 0 \\
	0 & 0 & xK\s_1-N\s_2 & L\s_1+M\s_2 \\
	-(xKM+LN)\s_2 &  A & RxL(L\s_1 + M\s_2)(y\PP_{xy} + \PP_x) & 0 \\
	x(xKM+LN)\s_1 & B & Rx^2M(L\s_1 + M\s_2)(y\PP_{xy} + \PP_x) & 0 
\end{pmatrix},
\]
where
\[
\begin{aligned}
&A=RyL(L\s_1 + M\s_2)(x\PP_{xx} + 2\PP_x)+ K(xK\s_1 - N\s_2),\\
&B=RyxL(L\s_1 + M\s_2)(x\PP_{xx} + 2\PP_x)-N(xK\s_1 - N\s_2).
\end{aligned}
\]
Its determinant has the form
\[
\det \smbl(E_q) = -k(KMx+LN)(L\s_1 + M\s_2)^2
\left( Rxy(L\s_1 + M\s_2)^2(x\PP_{xx} + 2\PP_x)+(xK\s_1 - N\s_2)^2 \right), 
\]
which gives us three characteristic vector fields of $E_q$:
\[
Z_1=L\Sdir{x}+M\Sdir{y},\qquad
Z_{2,3}=xK\Sdir{x}-N\Sdir{y} \pm \sqrt{-Rxy(x\PP_{xx} + 2\PP_x)} (L\Sdir{x}+M\Sdir{y}) ,
\]
where the two vector fields $Z_{2,3}$ exist when $x\PP_{xx} + 2\PP_x\leq 0$, which is always true for the properly defined thermodynamic states \cite{Lychagin2020}. 

Let us consider solutions of the quotient equation \eqref{eq:QuotientEa^2} that are constant along these characteristics vector fields. We will call this special class of solutions \textit{constant-type solutions.}

Note that for particular differential equations having constant-type solutions makes possible to find all solutions of a differential equation. For example, all solutions of the equation $u_{xy}=0$ (its characteristic fields are $\Sdir{x}$ and $\Sdir{y}$) are sums of constant-type solutions, i.~e., solutions of $\Sdir{x}u=0$ and $\Sdir{y}u=0$.

Of course, in our case this method does not lead to all solutions of the quotient equation, but, in this discussion, we restrict ourselves to constant-type solutions.

Firstly, we consider a characteristic vector field $Z_1=L\Sdir{x}+M\Sdir{y}$ and the ideal gas model, i.~e., when the Planck potential
\begin{equation*}
	\PP(x,y) = \frac{n}{2}\ln y - \ln x,
\end{equation*}
where $n$ is a number of freedom degrees of a gas particle. 

Let us require the functions $M$ and $L$ be the first integrals of the vector field $Z_1$. Then solving an overdetermined system $E_q\cup\{ Z_1(L)=Z_1(M)=0 \}$, we obtain the following of solutions of $E_q$

\begin{equation}\label{eq:quotsol1}
 	L=0,\; M=c_1x^{1+\frac{2}{n}},\; K=\sqrt{c_2x^2-\omega^2},\;  N=-\dfrac{2y\sqrt{c_2x^2-\omega^2}}{n},
\end{equation} 
\begin{equation}\label{eq:quotsol2}
L=c_1\left(\dfrac{x}{y}\right)^{\frac{2n}{n-2}},\;
	M= \dfrac{y}{x} L,\;
	K=\sqrt{c_2\left(\dfrac{x}{y}\right)^{\frac{2n}{n-2}}-\omega^2}, \; 	N=-\dfrac{2Ky}{n},
\end{equation}
where $c_1\neq 0$, $c_2$ are constants.

\section{Euler solutions}
In this section we use previously obtained solutions of the quotient equation to solve Euler system.

Consider the first solution \eqref{eq:quotsol1}
\[
L=0,\quad M=c_1x^{1+\frac{2}{n}},\quad K=\sqrt{c_2x^2-\omega^2},  \quad N=-\frac{2y\sqrt{c_2x^2-\omega^2}}{n}.
\]
Recalling that $x,y,K,L,M,N$ are Tresse coordinates and adding the obtained solutions to the Euler system \eqref{eq:Euler} for the case of ideal gas we get a finite-type system 
\begin{equation*}
	\left\{
	\begin{aligned}
		&\dens_a=0, \quad \temp_a=c_1\dens^{1+\frac{2}{n}},  \quad \vel_a = \sqrt{c_2\dens^2-\omega^2},\\
		& \temp_t+{ \vel\temp_a }=-\dfrac{2\temp}{n}  \sqrt{c_2\dens^2-\omega^2} , \\
		&\dens(\vel_t  + \vel\vel_a)+ R\dens\temp_a + \omega^2\dens a=0,\\
		&\dens_t + \dens\vel_a=0,\\
		&2k \temp_{aa}
		-{Rn\dens}(\temp_t+\vel\temp_a)+2R\temp\dens_t =0.
	\end{aligned}
	\right.
\end{equation*} 
This system is overdetermined and solving it we get
\[
\dens = \frac{\omega}{\sqrt{c_2}\cos(c_3-\omega t)}, \quad
\vel = a\,\omega \tan(c_3-\omega t)+f(t) ,
\]
\[
\temp = c_1 a \dens^{1+\frac{2}{n}} - \cos^{-\frac{2}{n}}(c_3-\omega t) 
\left(  c_1c_2^{-\frac{n+2}{2n}}\omega^{\frac{n+2}{n}} \int \frac{f(t)}{\cos(c_3-\omega t)}  dt +c_5 \right) ,
\]
where
\[
f(t) = \frac{-Rc_1c_2^{-\frac{n+2}{2n}}\omega^{\frac{n+2}{n}} }{\cos(c_3-\omega t)} \left( 
\int \cos^{-\frac{2}{n}}(c_3-\omega t) dt +c_4\right) 
\]
and  $c_1\neq 0, c_2>0,\ldots,c_5$ are  constants.

Doing the same with the solution \eqref{eq:quotsol2}
\[ 
L=c_1\left(\frac{x}{y}\right)^{\frac{2n}{n-2}}, \quad
M= L \frac{y}{x}  , \quad
K=\sqrt{c_2\left(\frac{x}{y}\right)^{\frac{2n}{n-2}} -\omega^2}, \quad 
N= \dfrac{-2Ky}{n},
\]
we have
\[
u=a\,\omega\tan(c_3-\omega t) +2f(t), \quad 
\temp = \dens \left(\frac{c_2\cos^2(c_3-\omega t)}{\omega^2} \right)^{\frac{n-2}{2n}} ,
\]
\[
\dens = \frac{c_1 \omega^2 a}{c_2\cos^2(c_3-\omega t)} - \frac{c_1\omega^2}{c_2\cos(c_3-\omega t)}\left( \int \frac{ 2f(t)}{\cos(c_3-\omega t)} dt+ c_5\right) ,
\]
where $f(t)$ is the same as before and $c_1,\ldots,c_5$ are constants.

Notice, the obtained solutions exist only on a certain time interval.    


\section{Virial expansion}\label{sec:virial}
In this section we find solutions of the quotient \eqref{eq:QuotientEa^2} in form of asymptotic expansion.

Let us consider the Planck potential $\PP$ in the form of \textit{virial expansion} \cite{LR}:
\[
\PP(x,y)=\frac{n}{2}\ln y - \ln x - \sum_{i=1}^{\infty}\frac{x^i}{i}A_i(y).
\]
Then we can find solutions of the system \eqref{eq:QuotientEa^2} in the form of power series of $x$:
\begin{align*}
	&K(x,y)=x^{d_K}\sum_{k=0}K_k(y)x^k,\quad &L(x,y)=x^{d_L}\sum_{k=0}L_k(y)x^k,\\
	&M(x,y)=x^{d_M}\sum_{k=0}M_k(y)x^k,\quad &N(x,y)=x^{d_N}\sum_{k=0}N_k(y)x^k,
\end{align*}
where $d_K$, $d_L$, $d_M$, $d_N$ are the integer constants that should be chosen such that the zeroth term of the expansion of the quotient equation \eqref{eq:QuotientEa^2} does not vanish. Choosing $d_K=0$, $d_L=1$, $d_M=0$, $d_N=0$ we can write the zeroth order term of this expansion:
\begin{equation}\label{eq:expansion0}
	\left\{
	\begin{aligned}
      	& M_0M_0^{\prime}=0, \\	
      	& M_0N_0^{\prime} - N_0M_0^{\prime}  -K_0M_0 =0, \\
		&(L_0N_0+M_0K_0)(N_0L_0^{\prime}+M_0K_0^{\prime}+K_0L_0)=0,\\
		&RM_0^2(y L_0^{\prime} +M_0^{\prime} + L_0  ) - L_0^{\prime}N_0^2 + M_0(K_0^2+\omega^2) -L_0K_0N_0=0.
	\end{aligned}
	\right.
\end{equation}
Keeping in mind that $M_0,L_0,K_0,N_0 \neq 0$, we solve the first three equations of \eqref{eq:expansion0}:
\[
M_0=c_1, \quad K_0=N_0^{\prime},\quad
 L_0 = \frac{-c_1N_0^{\prime} +c_2 }{N_0} , 
\]
where $c_1\neq 0$ and $c_2$ are arbitrary constants.

The latter leads to the equation
\begin{equation*}
N_0^{\prime} = \frac{c_1c_2Ry +(\omega^2 y -c_3)N_0  }{c_1^2Ry-N_0^2} ,
\end{equation*}
where $c_3$ is an arbitrary constant. With the following transformations:
\[
y \mapsto \frac{Rc_1^4}{c_2^2}\,y,\quad N_0 \mapsto \frac{Rc_1^3}{c_2}\,N_0,
\]
we reduce this equation to the following
\begin{equation} \label{eqq}
N_0^{\prime} = \frac{AyN_0 +B N_0 +y}{y-N_0^2},
\end{equation}
where
$A=\frac{c_1^2 \omega^2}{c_2^2}$, $B=-\frac{c_3}{c_1^2 R}$.

The last ordinary differential equation depends on the two constants $A$ and $B$, so does its solution $\mathcal{N}_0^{A,B}(y)$, which we call {\it flow-temperature function,} since $N_0(y)$ is the zeroth-order term of the material derivative $N(x,y)$ (see Section \ref{sec:Q}).

It is hard to solve \eqref{eqq} analytically, thus, we use numerical methods to demonstrate qualitatively different solutions of this equation as curves on the plane $(y,N_0)$. Note that the constants $A$ and $B$ define qualitative properties of $N_0$, i.~e., number of singular points, their positions and types.

\begin{figure}[h]
	\centering
	\begin{subfigure}[b]{0.40\textwidth}
		\includegraphics[width=\textwidth]{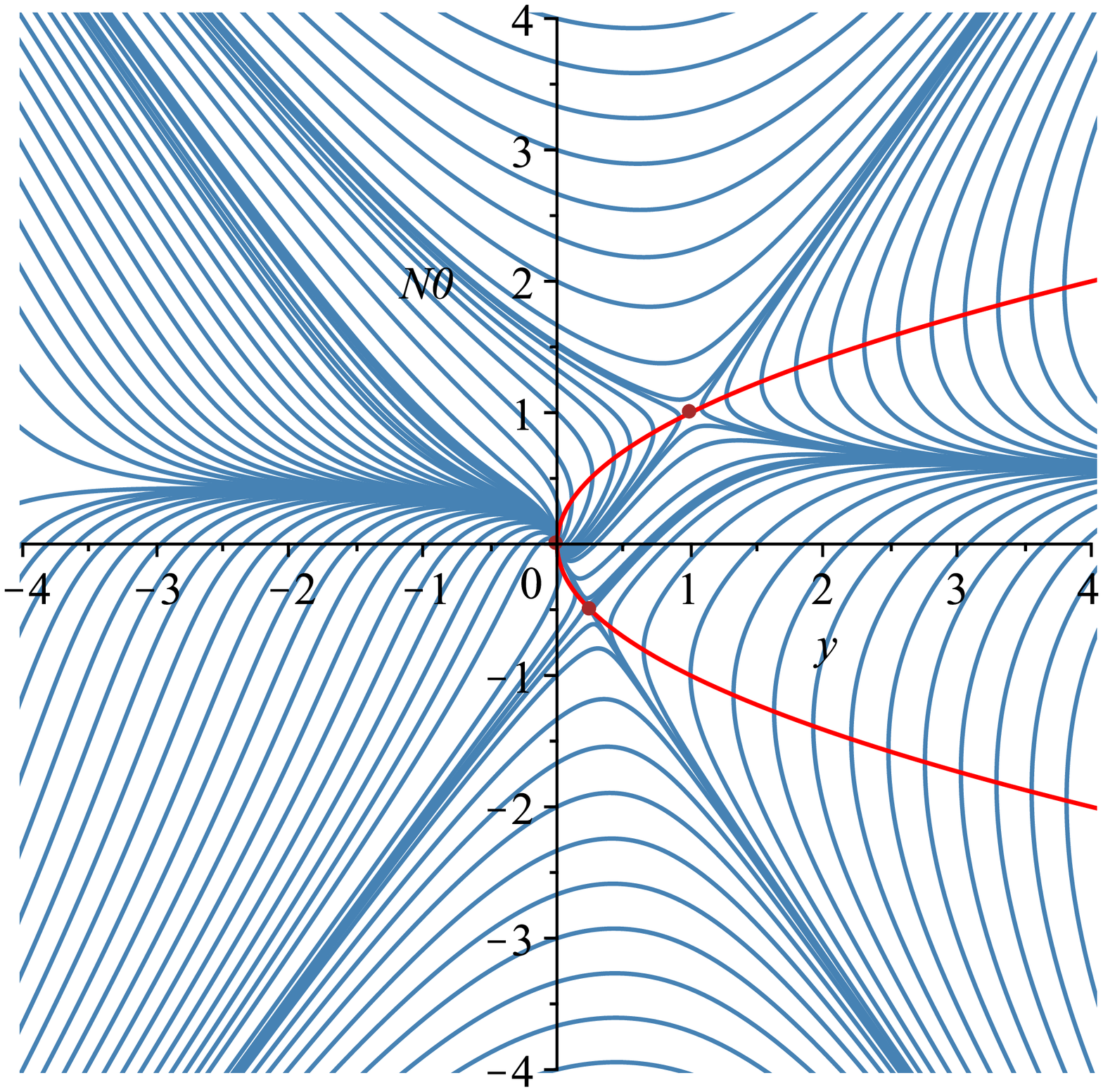}
		\caption{}\label{fig:pic1}
	\end{subfigure}  \qquad
	\begin{subfigure}[b]{0.40\textwidth}
		\centering
		\includegraphics[width=\textwidth]{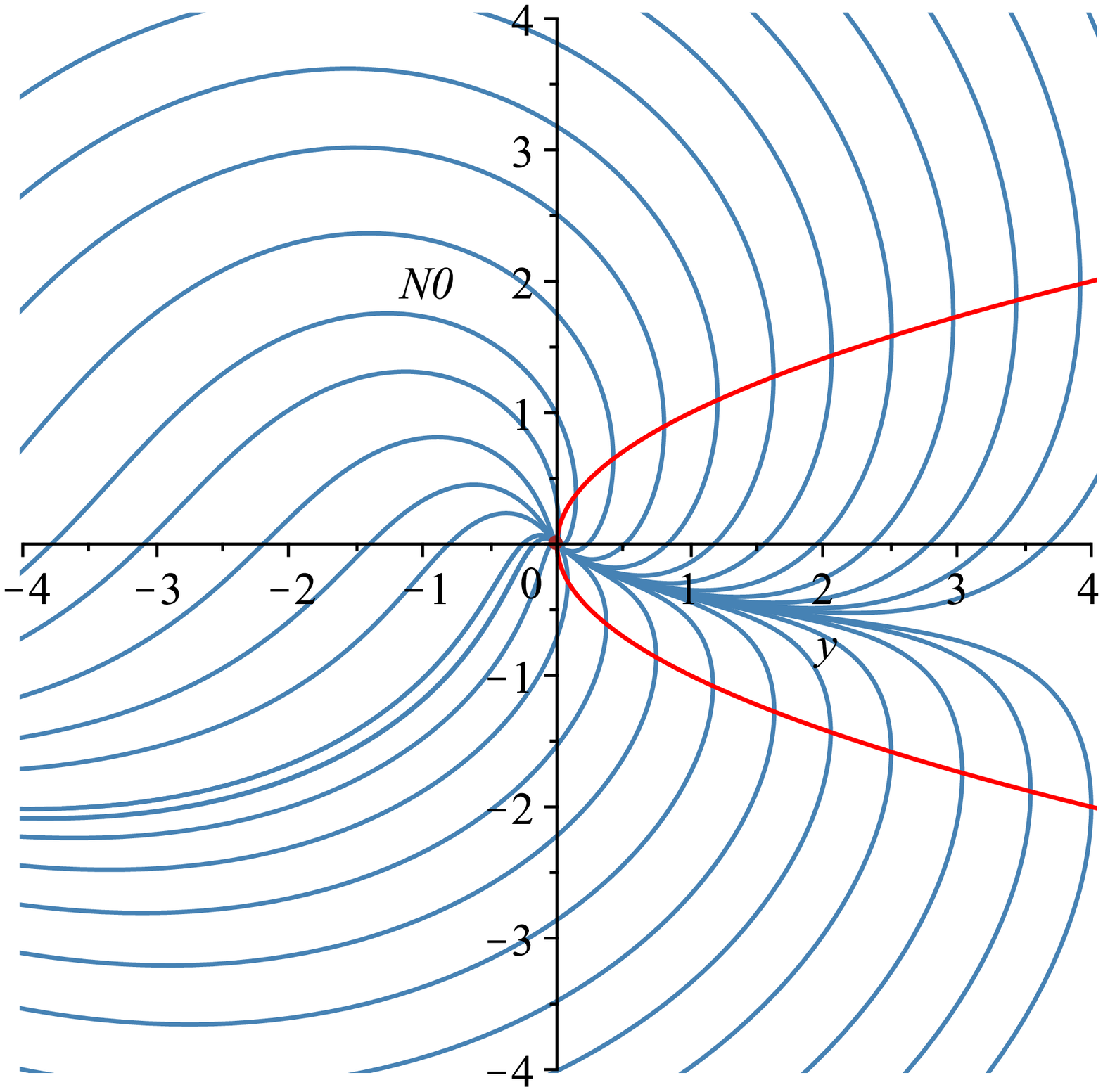}
		\caption{}\label{fig:pic2}
	\end{subfigure}
    \begin{subfigure}[b]{0.40\textwidth}
	    \includegraphics[width=\textwidth]{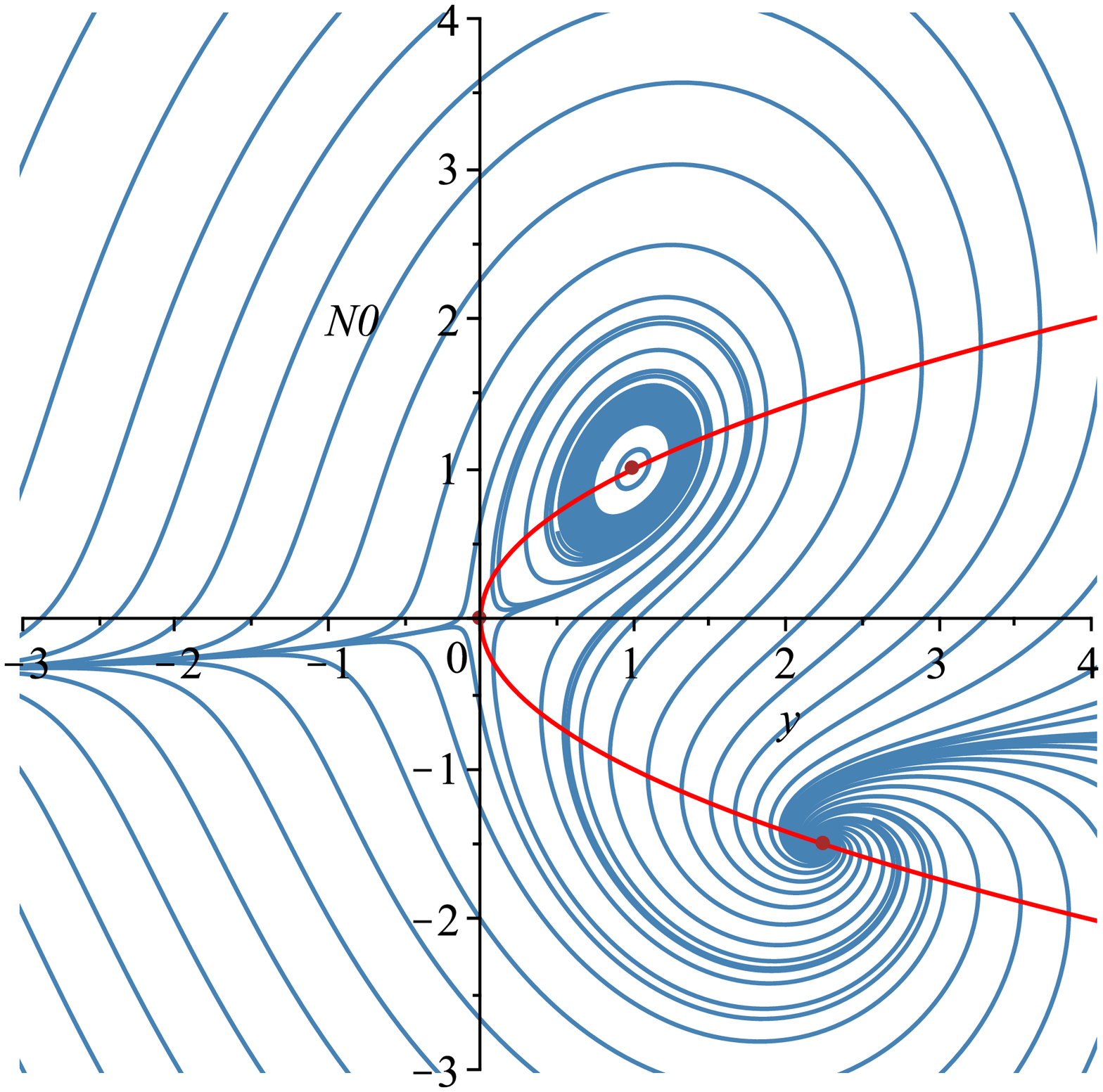}
		\caption{}\label{fig:pic3}
    \end{subfigure}  \qquad
	\begin{subfigure}[b]{0.40\textwidth}
		\centering
		\includegraphics[width=\textwidth]{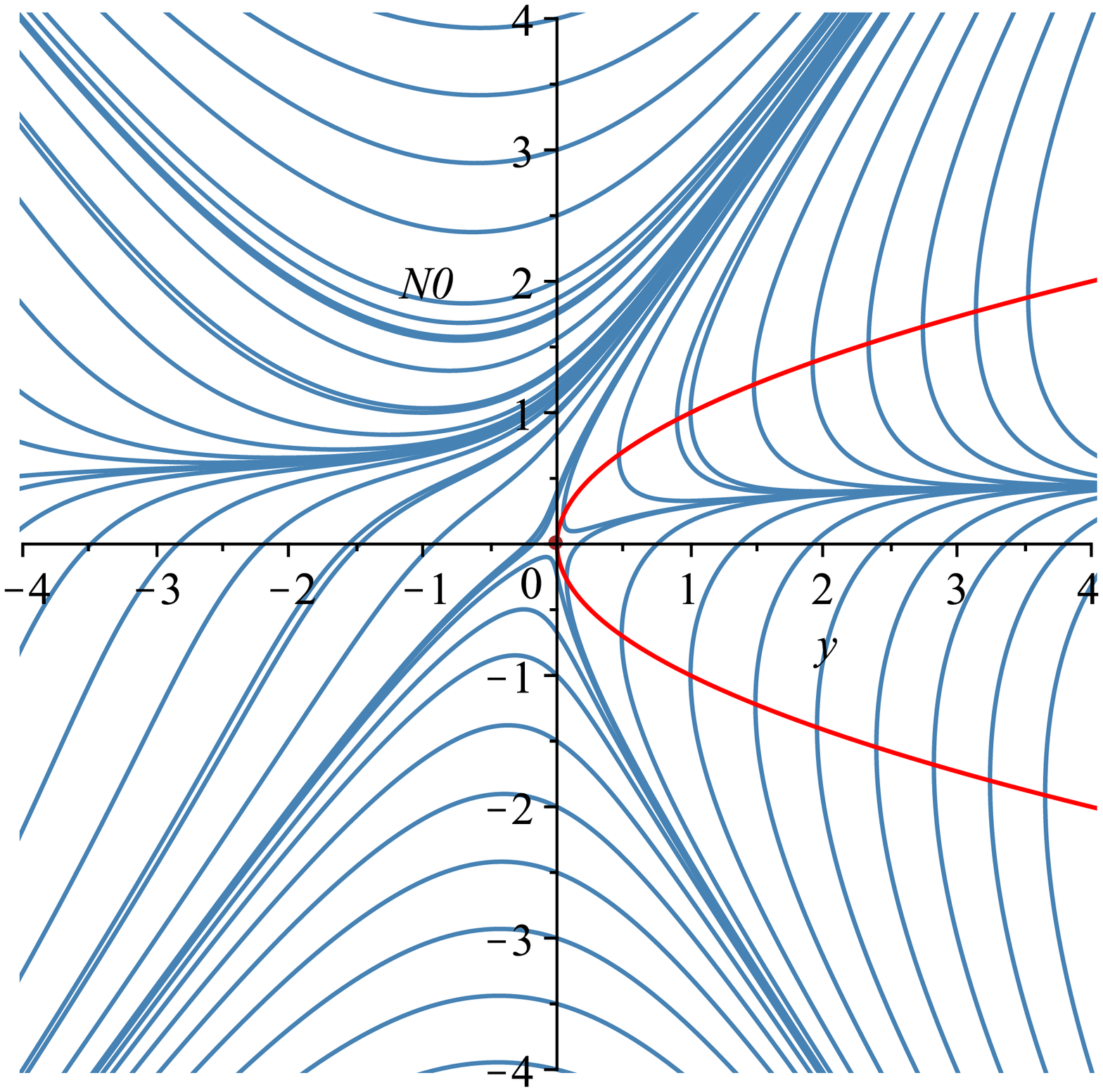}
		\caption{}\label{fig:pic4}
\end{subfigure}
	\caption{}
\end{figure}

The solution graphs for some specific values of $A$ and $B$ are given below. Note that the red parabola ($y=N_0^2$) in Figures \ref{fig:pic1}--\ref{fig:pic4} is the breaking points of the solutions.

\begin{itemize}
	\item Figure \ref{fig:pic1}. $A=-2$, $B=1$. Singular points: $(0,0)$ -- non-stable node, $(\frac{1}{4}, -\frac{1}{2})$, $(1,1)$ -- saddle points.
	\item Figure \ref{fig:pic2}. $A=1$, $B=2$. Singular point $(0,0)$ -- non-stable node.
	\item Figure \ref{fig:pic3}. $A=2$, $B=-3$. Singular points $(0,0)$ -- saddle point, $(1,1)$ -- centre point, $(\frac{9}{4}, -\frac{3}{2})$ -- non-stable spiral.  
	\item Figure \ref{fig:pic4}. $A=-2$, $B=-1$. Singular point $(0,0)$ -- saddle point.
\end{itemize}

\begin{figure}[h]
	\centering
	\begin{subfigure}[b]{0.40\textwidth}
		\includegraphics[width=\textwidth]{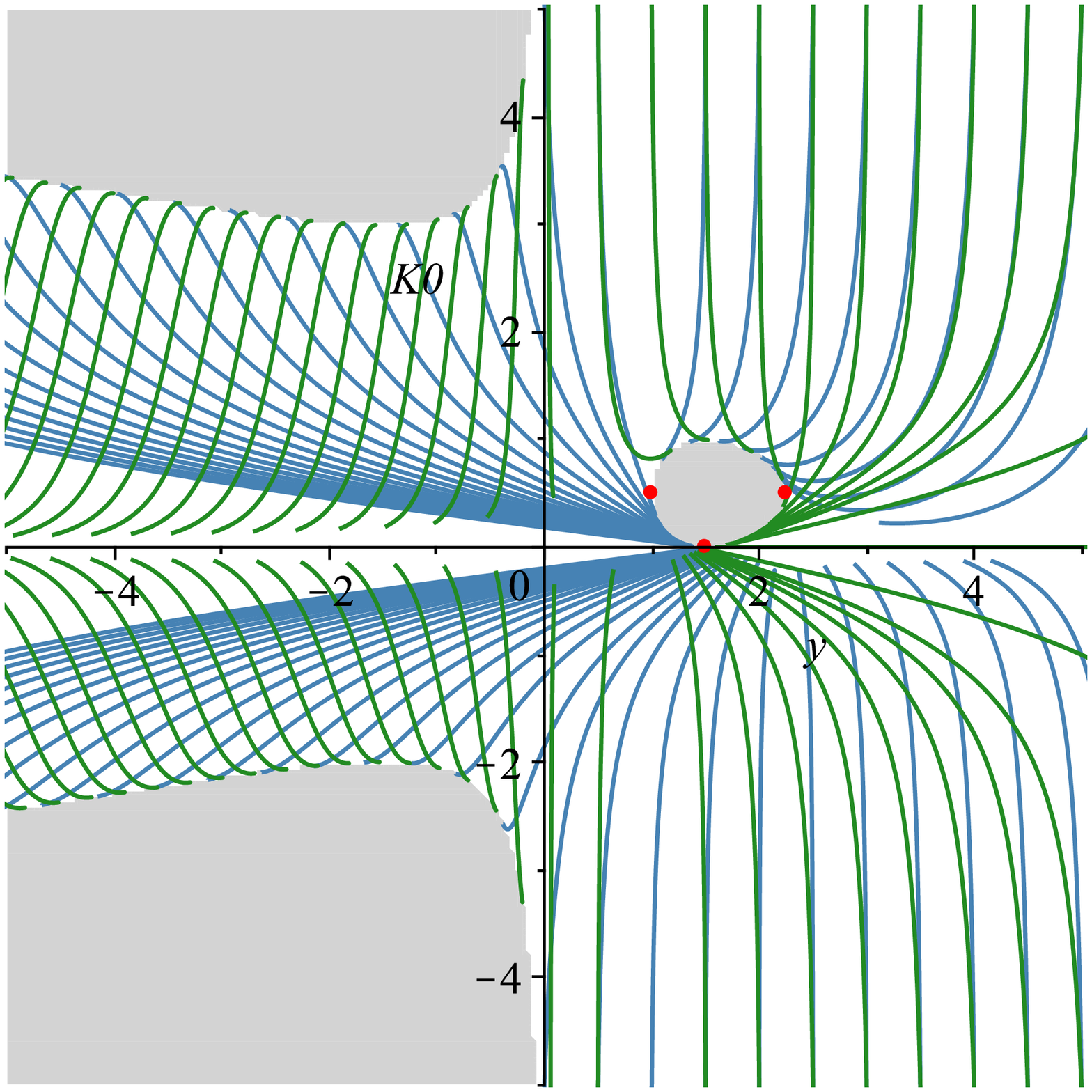}
		\caption{}\label{fig:picK}
	\end{subfigure}  \qquad
	\begin{subfigure}[b]{0.40\textwidth}
		\centering
		\includegraphics[width=\textwidth]{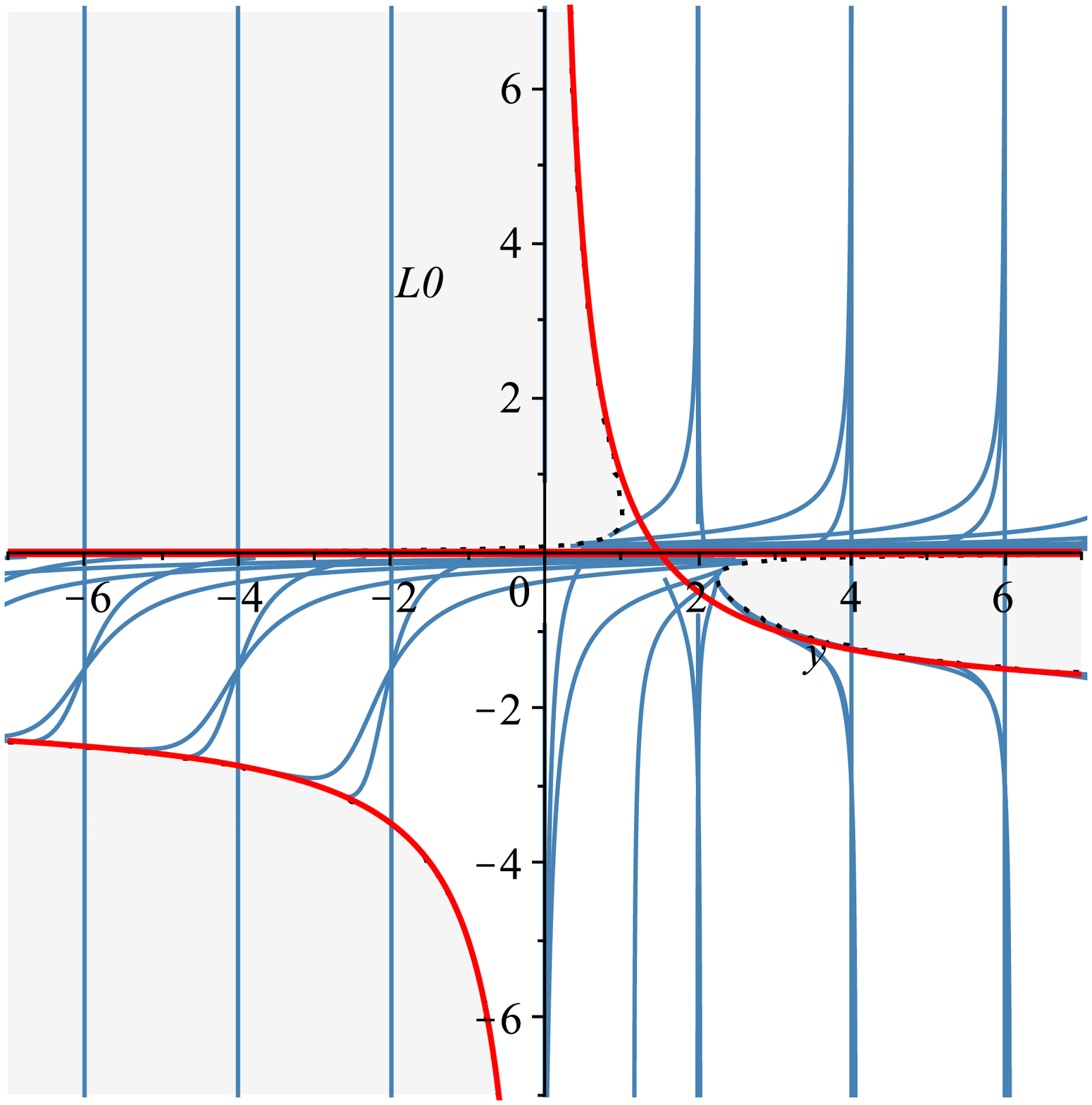}
		\caption{}\label{fig:picL}
	\end{subfigure}
	\caption{}
\end{figure}

In Figure \ref{fig:picK}, we show graphs of the function $K_0$ corresponding to the function $\mathcal{N}_0^{2,-3}(y)$ (Figure \ref{fig:pic3}). The green and blue curves are two different branches of a solution at each point of the plane besides the areas painted grey, where no solution exists. The three singular points are painted red.

Solutions for the function $L_0$ are even more complex. In the figure \ref{fig:picL}, we depict graphs of $L_0$ corresponding to the function $\mathcal{N}_0^{2,-3}(y)$ and $c_1=c_2=1$. We can see that the plane is divided into two parts. The first one, where only one solution exists, is of grey colour, and the other one, where through each point three solution pass, is white. The singular points lie on the red curves.

The ODE system for the first order terms of the expansion $M_1$, $L_1$, $K_1$ and $N_1$ is the following
\begin{equation}\label{eq:expansion1}
	\left\{
	\begin{aligned}
		&kM_0M_1^{\prime} +k(M_0^{\prime} + L_0)M_1-R (yK_0+\frac{nN_0}{2}) =0,\\
		&N_0M_1^{\prime} - M_0N_1^{\prime} + N_1(M_0^{\prime} - L_0) + K_1M_0 - M_1N_0^{\prime}=0,\\
		& \left( RM_0y - N_0^2\right) L_1^{\prime} + RM_0M_1^{\prime}
		+(M_0K_0 - 2L_0N_0)K_1 -
		(K_0L_0 + 2L_0^{\prime}N_0)N_1 +\\
		&+((2yL_0^{\prime} + 3L_0 + 2M_0^{\prime})RM_0 + K_0^2 + \omega^2)M_1 + RM_0(yL_0 + M_0)L_1 +\\
		&RM_0(M_0(M_0(A_1y )^{\prime\prime} + (yA_1)^{\prime} (3L_0 + M_0^{\prime})) + 2yA_1(L_0^2 + M_0L_0^{\prime}))=0,\\
		&(L_0N_0 + M_0K_0)K_1^{\prime} + (RyL_0M_0 + N_0K_0)L_1^{\prime} + RL_0M_0M_1^{\prime}+
		N_1(K_0^{\prime}L_0 + L_0^{\prime}K_0)+ \\
		&L_1(RM_0(yL_0^{\prime} + 2L_0 + M0^{\prime}) + yRL_0^2 + K_0^{\prime}N_0 + K0^2 + \omega^2)+\\ &M_1(RL_0(yL0^{\prime} + 2L_0 + M_0^{\prime}) + K_0^{\prime}K_0) + K_1(M_0K_0^{\prime} + 4K_0L_0 +L_0^{\prime}N_0) + \\
		&	RL_0(M_0(M_0(A_1y )^{\prime\prime} + (yA_1)^{\prime} (3L_0 + M_0^{\prime})) + 2yA_1(L_0^2 + M_0L_0^{\prime})) =0.
	\end{aligned}
	\right.
\end{equation}

Substituting functions $M_0$, $L_0$, $K_0$ and $N_0$ into the system \eqref{eq:expansion1}, we get the solution for the first order terms $M_1$, $L_1$, $K_1$ and $N_1$.

The ODE system for the second order terms of the expansion has the same type as \eqref{eq:expansion1}, i. e., it consists of four linear equations with the coefficients that depend on the zeroth and first order terms of the expansion.    
Similarly, we can obtain any number of terms in the expansion of the functions $M$, $L$, $K$ and $N$.

\textbf{Acknowledgments.} The research was partially supported by the Russian Science Foundation Grant 21-71-20034.

%
%
%
%
%

\end{document}